\documentclass[10pt,twocolumn,a4paper]{article}
\usepackage{times}
\usepackage{a4wide}
\usepackage[utf8]{inputenc}
\usepackage{latexsym,amsbsy,amsmath}
\usepackage[pdftex]{graphicx}
\usepackage{xcolor}
\usepackage{longtable}
\usepackage{isotope}
\usepackage{microtype}
\usepackage[bitstream-charter]{mathdesign}
 \DeclareGraphicsExtensions{.eps, .jpg,.png}
\renewcommand{\vec}[1]{\boldsymbol{#1}}

\def\ll#1#2{\tilde{\lambda}_{#1}.\tilde{\lambda}_{#2}}
\def\llss#1#2{\tilde{\lambda}_{#1}.\tilde{\lambda}_{#2}\,\boldsymbol{\sigma}_{#1}\boldsymbol{\sigma}_{#2}}
\begin{document}
\title{String dynamics and metastability of fully-heavy  tetraquarks }
\date{March 2nd, 2017}
\author{Jean-Marc~Richard\footnote{%
email: j-m.richard@ipnl.in2p3.fr}
\\
{\small Universit\'e de Lyon, Institut de Physique Nucl\'eaire de Lyon,
IN2P3-CNRS--UCBL,}\\
{\small 4 rue Enrico Fermi, 69622  Villeurbanne, France}
\\[2pt]
A. Valcarce\footnote{%
email: valcarce@usal.es}
\\
{\small Departamento de F{\'\i}sica Fundamental and IUFFyM,}
{\small Universidad de Salamanca, 37008 Salamanca, Spain}
\\[2pt]
J.~Vijande\footnote{%
email: javier.vijande@uv.es}
\\
{\small Unidad Mixta de Investigación en Radiofísica e Instrumentación Nuclear en Medicina (IRIMED),}
 \\
 {\small Instituto de Investigación Sanitaria La Fe (IIS-La Fe)}
 \\
{\small Universitat de Valencia (UV) and IFIC (UV-CSIC), Valencia, Spain}}
\onecolumn
\maketitle
\begin{abstract}
Multiquark states have been advocated to explain recent experimental data in the heavy-light sector, and there are already speculations about multiquarks containing only heavy quarks and antiquarks. With a rigorous treatment of the four-body problem in current quark models, full-charm  $(cc\bar c\bar c)$ and full-beauty $(bb\bar b\bar b)$ tetraquarks are found to be  unbound. Thus their stability should rely on more subtle effects that are not included in the simple picture of constituent quarks. The case of $(bc\bar b\bar c)$ might be more favorable if the naive color-additive model of confinement is replaced by a string-inspired interaction.
\end{abstract}
%%%%
\twocolumn
\section{Introduction}
\label{se:intro}
Recent discoveries of new particles hint at combination of quarks in a way not seen before: pentaquarks \cite{2015Natur.523R.267C}, and four-quark states \cite{LHCB-CC}, namely particles with hidden heavy flavor and a pair of light quarks, $(Q\bar Q q\bar q)$. However, in this sector, as well as for other multiquark states like dibaryons or pentaquarks,  the experimental candidates are resonances, lying above their dissociation threshold. These resonances have  been discussed in many papers, but in theoretical studies, some special attention is also paid for configurations which would be stable against spontaneous breaking, or at least metastable, i.e., lying below their nearest threshold. 

In particular, several recent papers speculate on the existence of tetraquarks made of four heavy constituents, $(QQ\bar Q\bar Q)$, either in the charm or in the beauty sector \cite{Karliner:2016zzc,Bai:2016int,Wu:2016vtq,Wang:2017jtz}. Earlier papers on the subject include \cite{Iwasaki01081975,Carlson:1991zt,Lloyd:2003yc}, (the first one as early as 1975 !), as well as papers in which the  equal-mass case is compared to other flavor configurations \cite{Zouzou:1986qh, Vijande:2009kj}. These studies are rather timely, as the evidence for several $XYZ$ states has demonstrated that hadron and electron colliders provide good opportunities to extend our knowledge of heavy-flavor spectroscopy. Experimentally, states with hidden heavy flavor offer some advantages, as they can be detected with the help of triggers such as $J/\psi$ which are very efficient. Other configurations are seemingly more delicate, as illustrated by the difficulties encountered in the search for double-charm baryons and double-charm tetraquarks.

On the theory side, binding a multiquark configuration is not as obvious as it may look at first sight. For instance, Lipkin (private communications and, e.g., \cite{Lipkin:1986dw,Lipkin:2007cg}) used to stress that for a set of two mesons at rest there are only two three-dimensional kinetic-energy operators, while a tetraquark involves three of them.\footnote{The argument was given for pedagogical purpose. Lipkin was of course fully aware that, thanks to the virial theorem, the kinetic energy tends to readjust itself independently of the number of operators.}\@ So, one should get some good dynamical effect to overcome this handicap. In the case of atomic physics, the situation is also rather delicate: in 1945, Wheeler speculated about the existence of the positronium molecule $\mathrm{Ps}_2=(e^+,e^+,e^-,e^-)$ (the paper was published in 1946 \cite{1946NYASA..48..219W}); a first calculation by Ore concluded that this four-electron configuration is likely unstable \cite{PhysRev.70.90}; but the following year, the very same Ore, associated with Hylleraas, published an elegant analytic proof of the stability \cite{PhysRev.71.493}. In fact, the stability or instability of few-charge systems depends rather critically of the masses which are involved, see, e.g., \cite{ARV}.  In particular, the very tiny binding of Wheeler's positronium molecule contrasts with the comfortable binding of the hydrogen molecule $\mathrm{H}_2$. 

The constituent quark model follows rather closely the patterns of few-charge systems, when the dynamics is taken as an additive flavor-independent and spin-independent interaction, see Sec.~\ref{se:CEM}. The chromomagnetic interaction offers some opportunities for multiquarks, that will be briefly reviewed in Sec.~\ref{se:CMI}. Another improvement comes from the string model, which suggests a multi-body variant of the linear part of the chromoelectric potential, which gives more attraction, provided the quarks (and the antiquarks) are not constrained by Fermi statistics. This gives $(bc\bar b\bar c)$ some opportunities that equal-mass configurations such as $(bb\bar b\bar b)$, $(cc \bar c\bar c)$ do not share. The case of $(bb\bar c\bar c)$ or c.c.\ is more delicate, as it benefits from $C$-conjugation breaking, which makes $\mathrm{H}_2$ more stable than $\mathrm{Ps}_2$, but is submitted to the constraints of the Fermi statistics. 

In this article, we wish to extend some of the existing results on fully-heavy tetraquark states, and explain why their binding depends on their detailed flavor content. We also address the question of metastability, which is also relevant for the $XYZ$ states. Many scenarios lead to $(bc\bar b\bar c)$ that is below the $(b\bar c)+(c \bar b)$ threshold, but leave the decay into bottomonium and charmonium permitted. This is similar to the question of metastability of the hydrogen-antihydrogen molecule \cite{2004FBS....34...63F}.

\section{Chromoelectric model}\label{se:CEM}
\subsection{Color-additive model}
In the limit of very heavy quarks, the chromomagnetic forces vanishes. It is thus interesting to consider the case of a purely chromoelectric interaction
\begin{equation}
 \label{eq:colour-add}
 H=\sum_i \frac{{\vec p}_i^2}{2\,m_i}-\text{c.o.m.}  +V~,\quad
 V=-\frac{16}{3}\sum_{i<j} \ll{i}{j}\,v(r_{ij})~.
\end{equation}
This is of course a very crude modeling, with non-relativistic kinematics, two-body forces, color treated as a global operator, etc., but, at least, it can be used as benchmark. Here, $ \ll{i}{j} $ should be suitably modified for quark-antiquark pairs. The normalization is such that $v(r)$ is the quarkonium potential, something like $v(r)=-a/r+\lambda\,r$ à la Cornell~\cite{Eic75}, or $v(r)=A\,\ln(r/c)$ à la Quigg and Rosner~\cite{Quigg:1979vr}, or $v(r)=A_,r^\alpha+B$ à la Martin~\cite{Martin:1980rm}.  

The latter choice was adopted by Zouzou et al.~\cite{Zouzou:1986qh} who studied $(QQ\bar q\bar q)$ as a function of the quark mass ratio $M/m$ and found that a pretty large $M/m$ is needed to achieve stability below the threshold for decay into two flavored mesons. With current quark models, stability is achieved with $(bb\bar u\bar d$). The stability of $(cc\bar u\bar d)$, first obtained by Janc and Rosina~\cite{Janc:2004qn}, and confirmed in further work \cite{Vijande:2007rf}, makes use of a favorable chromomagnetic interaction, and would require a larger $M/m$ if the chromomagnetic term is removed. Thus the result of Llyod and Vary \cite{Lloyd:2003yc}, claiming for the existence of a stable $(cc\bar c\bar c)$,  was received with some skepticism. 

The stability of $(QQ\bar q\bar q)$ in the limit of large quark-to-antiquark mass ratio, as a consequence of flavor independence, is nowadays an old prediction (about 35 years), but it has never been tested! The physics of double charm is seemingly difficult, even with modern detectors. The existence of the simpler $(QQq)$ configuration, i.e., doubly charmed baryons is not even established  \cite{Olive:2016xmw}. 

\subsection{A mathematical digression}

The analogy between the stability of few-charge systems and multiquarks in additive chromoelectric potential offers a good guidance for identifying the favorable configurations. There are, however, some differences, not so much due to the radial shape of the potential, but mainly due to the color algebra replacing the simpler algebra of electric charges. 

In refs.~\cite{Richard:2010jh,Richard:2016eis}, there is an attempt to explain why, unlike in the case of the positronium molecule, the equal-mass systems are unstable in the chromoelectric model with frozen color wave functions. In both the atom and quark cases, the four-body system and its threshold, after simple rescaling, are governed by a generic Hamiltonian
\begin{equation}
 \label{eq:H4}
 H=\sum_i {\vec p}_i^2-\text{c.o.m.}  +\sum_{i<j} g_{ij}\,v(r_{ij})~.\qquad \sum_{i<j} g_{ij}=2~,
\end{equation}
with $v(r)=-1/r$ in the atomic case, and the quarkonium potential in the hadron case. Of course, all such Hamiltonians give a ground-state near the two-atom or two-meson threshold.  A symmetric distribution of the couplings, $g_{ij}=1/3~\forall i,j$, gives the largest energy, and any asymmetry in $\{g_{ij}\}$  lowers its energy. And one realizes that the algebra of color is less favorable than the algebra of charge-products in molecules, namely that  $\{g_{ij}\}$ is less asymmetric for a tetraquark than for the positronium molecule. Or say, multiquark spectroscopy is penalized by the non-Abelian nature of color.

It is important to stress that the color configurations, though they give the same cumulated strength $\sum g_{ij}$, are \emph{not} equivalent for the confining energy. Thus a color configuration which is potentially favorable for the chromomagnetic interaction (to be discussed below) might be far from optimal for the chromoelectric one~\cite{Vij16}. 

The above reasoning on the ground state of \eqref{eq:colour-add} as a function of $\{g_{ij}\}$ holds for a single color channel.  It is observed in explicit computations than the mixing of color states does no help much. 
\subsection{Solving the four-body problem}
To compute the ground-state of a three-body baryon in the quark model, a crude variational approximation is often sufficient. The accuracy is not very crucial given the crudeness of the model, and the exact wave function has a simple structure. The situation would be even better for a $N$-body baryon in the large-$N$ extension of QCD.

For a tetraquark close to its threshold, this is drastically different.  One has to estimate precisely $(q_1q_2\bar q_3\bar q_4)$ and its thresholds, to see whether there is a bound state. Moreover,  the $(q_1q_2\bar q_3\bar q_4)$ wave function has a $[(q_1\bar q_3)(q_2\bar q_4)]$ component and  a $[(q_1\bar q_4)(q_2\bar q_3)]$ one, corresponding to its `molecular' part, perhaps a $[(q_1 q_2)(\bar q_3\bar q_4)]$ diquark-antidiquark component, and a collective component that prevails in the event of deep binding. This has been discussed in detail in Ref.~\cite{Vij09}. A similar situation is encountered in atomic physics: the deeply bound He atom $(\alpha, e^-,e^-)$ is well described by a simple product of two  functions. For the weakly bound H$^-$ $(p,e^-,e^-)$, one has to introduce a much more subtle wavefunction to demonstrate the stability against dissociation into $\mathrm{H}+e^-$~\cite{ARV}.

More precisely, a simple Gaussian trial wave function
\begin{equation}
 \label{eq:G}
 \exp[-a (\vec x^2+\vec y^2 +\vec z^2)/2]=\exp\left[-a \sum \vec r_{ij}^2/4\right]~,
\end{equation}
where $\vec x$, $\vec y$ and $\vec z$ are properly normalized Jacobi variables describing the internal motion, would not distinguish among the various coupling distributions $\{g_{ij}\}$ envisaged in \eqref{eq:H4}. Nor will  a wavefunction assumed to be a function of $\vec x^2+\vec y^2 +\vec z^2$ only, which in the technical language of the four-body problem, is named the hyperscalar approximation.  

In practice, solving the toy model \eqref{eq:H4} or the four-quark problem in constituent models requires sophisticated tools. 
In Ref.~\cite{Bar07} the hyperspherical harmonic (HH) expansion was  pushed up to deal with systems of two quarks and two antiquarks with the 
same flavor, $(QQ\bar Q\bar Q)$. It was later on generalized to consider pairs of quarks with different flavor and  masses, $(Q\bar Q q \bar q)$ and $(QQ \bar q\bar q)$, where $q$ stands for a light quark~\cite{Vijande:2007rf}. Another approach is based on correlated Gaussians, in which the single wavefunction \eqref{eq:G} is replaced by a sum of correlated Gaussians
\begin{equation}
 \label{eq:Gp}
 \sum_n \alpha_n\left[ \exp[- \sum a_{ij}^{(n)}\, \vec r_{ij}^2/4]+\cdots]\right]~,
\end{equation}
where the ellipse are meant for terms deduced by symmetry. 
% The main difficulty of this method is to construct HH functions
% of proper symmetry for a system of identical particles~\cite{Bar98}. 
% In the HH formalism the three Jacobi vectors 
% are transformed into a single length variable,
% $\rho=(x^2+y^2+z^2)^{1/2}$, and $8$-angular variables, $\Omega$, that
% represent the location on the $8$-dimensional sphere. 
% The spatial basis states are given by 
% \begin{equation}
% \bra \rho \Omega |R\ket = U_{n}(\rho){\cal Y}_{ [K]} (\Omega) \, , 
% \end{equation}
% were ${\cal Y}_{ [K]} $ are the HH functions, and 
% $K$ is the grand angular momentum. 
% The Laguerre functions are used as the hyperradial basis
% functions $U_n(\rho)$.
% 
% A detailed discussion on the construction of color-spin states with 
% well-defined permutational symmetry as well as the
% SU(3) colour symmetry and charge conjugation $C$ has been presented in Ref.~\cite{Bar07}.
%
The Pauli principle leads to the restrictions
on the allowed combinations of spin-color-orbital basis states contributing to the $(Q'Q'\bar Q\bar Q)$
wave function. 
%
% \begin{itemize}
% \item[{(i)}] $(-1)^{S_{12}+\ell_1}=+1$, $(-1)^{S_{34}+I+\ell_3}=-1$ 
% for the $| 6_{12} \bar 6_{34} \ket$ color state,
% \item[{(ii)}] $(-1)^{S_{12}+\ell_1}=-1$, $(-1)^{S_{34}+I+\ell_3}=+1$ for 
% the $| \bar 3_{12} 3_{34} \ket$ state.
% \end{itemize}
In the $(Q \bar Q  Q' {\bar Q}')$ case, $C$-conjugation selects the spin-color-orbital configurations that can be combined in the  wave function.

Note that the dynamics of $(Q'Q'\bar Q\bar Q$) is simpler, with a single threshold. In the case of $(Q Q'\bar Q {\bar Q}')$ with $Q\neq Q'$, the lowest threshold is
$(Q\bar Q)+(Q'{\bar Q}')$ for a flavor-independent interaction, according to a theorem by Nussinov, Bertlmann and Martin \cite{Nussinov:1999sx,Bertlmann:1979zs}, and this remains often true when flavor-independence is broken by a chromomagnetic term. This means for, e.g., $(c\bar c q \bar q)$ configurations tentatively describing the $XYZ$ states, that one has to work in the continuum, without guaranty that a bound-state approximation is justified. 

% on the $Q \bar Q  n \bar n$ system as well.
% Coupling the color states of a quark and an antiquark can yield two possible
% representations: the singlet and the octet.
% These representations should be combined in the following way
% $\{|1_{12}1_{34}\ket, | 8_{12}, 8_{34}\ket \}$ to yield a total color singlet state~\cite{Jaffe:1976yi}.
% However, these states have not definite symmetry under particle
% permutations $(12)$ and $(34)$.
% To construct symmetrized states for the $Q\bar Q$ pair we
% consider the following combinations,
% \begin{equation}
%  | C_{12}^{\Gamma_{12}} \ket = \frac{1}{\sqrt 2} ( | C_{12} \ket + \Gamma_{12}
%    | C_{21}\ket  \, ,
% \end{equation}
% were $C_{12}=\{1_{12},8_{12}\}$, and $\Gamma_{12}=+1$ for a symmetric
% combination and $-1$ for an antisymmetric one. For light quarks the
% color and isospin states should be combined together to form states with well
% defined symmetry. The total color-isospin states are not only
% good symmetry states but also good $C-$parity states with,
% $C=\Gamma_{12}\Gamma_{34}$. Imposing the Pauli principle for the $Q\bar Q n
% \bar n$ system we get the following restrictions:
% $\Gamma_{12}(-1)^{S_{12}+\ell_1}=+1$, $\Gamma_{34}(-1)^{S_{34}+\ell_3}=+1$, on
% the basis states.

The four-body problem is notoriously delicate, as illustrated by the difficulties encountered and eventually overcome by Ore for the positronium molecule. Approximations are thus welcome, especially if they point out the main degrees of freedom. For instance the hydrogen molecule is well understood within the Born-Oppenheimer approximation, see, e.g., \cite{bransden2003physics}. The Born-Oppenheimer method translates in the quantum domain standard approximations of classical physics based on the differences of time scales: sequential radioactivity following primary fusion, melting of a ice sphere within a large vessel, or spontaneous penetration of a horseshoe in the ice \cite{reif2009fundamentals}, for which a quasi-equilibrium state is assumed at any time.  The Born-Oppenheimer method has been applied to $(QQq)$ and  heavy hybrids, and has been generalized to $XYZ$ states, see, e.g., \cite{Braaten:2014qka}.

Some other approximations imply a redefinition of the dynamics -- which might be justified from elaborate QCD studies --, but are not direct consequences of the simple models such as \eqref{eq:colour-add}. This is notoriously the case for the diquark model. Take for instance the harmonic-oscillator model of baryons. With standard Jacobi variables, it reads
\begin{equation}
 \label{eq:HO1} H=(\vec p_x^2+\vec p_y^2)/m + 3\,K\, (\vec x^2+\vec y^2)/2~,
\end{equation}
with a factor coming from $r_{12}^2+r_{23}^2+r_{31}^2=3\,(\vec x^2+\vec y^2)/2$. In the naive diquark approximation, one solves first the problem for $(1,2)$ alone, and then for $[(1,2),3]$, and one misses a factor $(3/2)^{1/2}$ for the part of the ground-state energy associated with $\vec x$.  The effect is not completely negligible, and is \emph{antivariational}. 

If one repeats the same exercise with an equal-mass tetraquark and a color $\bar 3 3$ wave function, one gets
\begin{equation}
 \label{eq:HO2} H=(\vec p_x^2+\vec p_y^2+\vec p_z^2)/m + 3\,K\, (\vec x^2+\vec y^2)/4+ K\,\vec z^2/2~,
\end{equation}
and, again, the approximation is antivariational, with $(3/4)^{1/2}\to 1/2$ for each diquark, and the open question of whether this becomes worse for a non-harmonic interaction. 

In \cite{Karliner:2016zzc}, there is an interesting  statement that the  binding energy of a $QQ$ diquark of color $\bar3$ is half that of a $Q\bar Q$ singlet: the change in a $Q\bar Q$ potential $V=g\,r^\alpha$ from $g$ to $g/2$ results in a rescaling by a factor $2^{-2/(\alpha+2)}$, according to the seminal paper~\cite{Quigg:1979vr}. This means a reasonable logarithmic regime $\alpha\to 0$. On the other hand, for the diquark-antidiquark binding, the mass dependence is found $\propto m^p$, with $p=0.712$, which if identified with the behavior $m^{\alpha/(\alpha+2)}$ of a power-law interaction, suggests a nearly Coulombic regime. The mass increase from the quark-quark case to the diquark-antidiquark one does not justify such a change of regime.

% Another delicate issue deals with estimating the short-range correlation $C_{ij}=\langle \delta^{(3)}(\vec r_{ij}) \rangle$ which enters the chromomagnetic corrections, and some contributions to weak decays. When one started serious studies of parity-violating effects among electrons in atoms, it was realized that the wave functions available at that time were not good enough, and one had to either redo the calculation of atomic structures, or  use some tricks to speed up the convergence, see, e.g., \cite{1978PhRvA..18.2399H}. 

% If one estimates the ground-state of a 2-body system bound by a linear interaction using a single Gaussian as trial function, and computes the corresponding $C_{12}$, one finds only $\sim 85\%$ of the correct value, but with a Coulomb interaction, this is only $24\%$! \footnote{Of course, the Schwinger rule~\cite{Quigg:1979vr,1978PhRvA..18.2399H} would improve the accuracy, that is to say $C_{12}$ computed as $\int v'(r) u^2(r) \mathrm{d}r$, where $u$ is the reduced radial wave function, instead of $C_12$ being read from $u(r)$ as $r\to 0$}\@
% One may always use an ad-hoc prescription for $C_{ij}$, for instance adopt the value fitted in another hadron, but then the model becomes more empirical!

Another difficulty arise when adding to the Hamiltonian a regularized form of the spin-spin interaction. Then, solving accurately the few-body problem  with a  superposition of linear confinement and short-range  terms becomes rather delicate.\footnote{We thank Emiko Hiyama for discussions on this point and many other topics.}

To sum up, the technical aspects of multiquark spectroscopy should not be underestimated, even if they are less exciting and challenging than the question of the underlying dynamics. For instance, years ago, Semay and Silvestre-Brac designed an empirical potential model, sometimes referred  to as AL1, to describe heavy and light hadrons. They then solved the $(cc\bar u\bar d)$ problem to look at possible bound states, and found no binding \cite{Semay:1994ht}, using an expansion on the eigenstates of a neighboring harmonic oscillator, a method which was widely used in nuclear physics. The very same potential was used later by Janc and Rosina \cite{Janc:2004qn} and Barnea et al.~\cite{Bar07} who found a $1^{+}$ state below the $DD^*$ threshold. 
The binding is expected to increase for the $(bb\bar u\bar d)$ system.  But for its hidden-beauty analogue $(b \bar b q\bar q)$,  the competition between the two thresholds discussed above
would not drive binding for the beauty partner of the $X(3872)$~\cite{Car12}.

\subsection{Improved chromoelectric models}
\begin{figure}[h!]
 \centering
 \includegraphics[width=.1\textwidth]{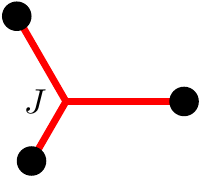}
 % strings-fig2.pdf: 0x0 pixel, 300dpi, 0.00x0.00 cm, bb=
 \caption{$Y$-shape interaction for the confinement of three quarks}
 \label{fig:Y}
\end{figure}

The color-additive interaction \eqref{eq:colour-add} can be improved to account for mechanisms suggested by non-perturbative QCD, in particular lattice simulations. In the case of baryons, Dosch et al.~\cite{Dosch:1975gf}, and many others, proposed to replace the so-called ''1/2'' rule, where the linear term $\lambda\,r$ of mesons becomes 
\begin{equation}\label{eq:one-half}
 V_{1/2}=\lambda\,(r_{12}+r_{23}+r_{31})/2~,
\end{equation}
by the $Y$-shape interaction (see Fig.~\ref{fig:Y}),
\begin{equation}\label{eq:Y}
 V_Y=\lambda\,\min_J(r_{1J}+r_{2J}+r_{3J})~. 
\end{equation}
As $V_Y\gtrsim V_{1/2}$~\cite{Dosch:1975gf}, the change $V_{1/2}\to V_Y$ pushes up the masses, but this is hidden by other uncertainties in the quark model of baryons.  

\begin{figure}[h!]
 \centering
 \includegraphics[width=.12\textwidth]{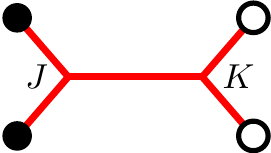}
 % strings-fig2.pdf: 0x0 pixel, 300dpi, 0.00x0.00 cm, bb=
 \caption{Double $Y$-shape, or ``butterfly'' interaction for the confinement of two quarks and two antiquarks}
 \label{fig:YY}
\end{figure}

In the case of tetraquarks, the string-inspired linear confinement is not always a repulsive correction as compared to the color-additive model. A connected contribution, sometimes named ``butterfly'' diagram (see Fig.~\ref{fig:YY}),%
\begin{equation}\label{eq:YY}
V_{YY}=\lambda\,\min_{J,K}(r_{1J}+r_{2J}+r_{JK}+r_{3K}+r_{4K})~,
\end{equation}
is obviously favorable if the two quarks are far away of the two antiquarks, as it merges two strings into a single one. 

\begin{figure}[h!]
 \centering
 \includegraphics[width=.12\textwidth]{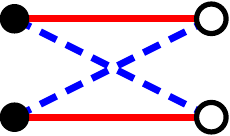}
 % strings-fig2.pdf: 0x0 pixel, 300dpi, 0.00x0.00 cm, bb=
 \caption{Flip-flop model for the confinement of two quarks and two antiquarks}
 \label{fig:FF}
\end{figure}

But the most dramatic effect comes from the ``flip-flop'' term  (see Fig.~\ref{fig:FF}).
\begin{equation}\label{eq:FF}
 V_\text{FF}=\lambda\,\min (r_{13}+r_{24},r_{14}+r_{23})~. 
\end{equation}
A precursor with quadratic confinement was introduced by Lenz et al.~\cite{Lenz:1985jk} for the study of meson-meson scattering. Then the model, in its linear version, was applied to tetraquarks and other multiquarks, with a good surprise and a warning \cite{Vijande:2007ix}.

The good surprise is that this interaction provides with more attraction, and enables to bind mass configurations $(m_1,m_2,m_3,m_4)$ that are left unbound by the color-additive model. 

The warning is that, when the tetraquark confinement is estimated as 
\begin{equation}\label{eq:tetraquark}
 V=\min(V_{YY},\;\lambda\,(r_{13}+r_{24}),\;\lambda\,(r_{14}+r_{23}))~,
\end{equation}
each term is associated with a different color wave function. It is a kind of Born-Oppenheimer effective interaction where the electrons of atomic physics are replaced by the gluon field, and the nuclei by the heavy quarks and antiquarks. For identical quarks, restrictions apply. When one attempts to restore the proper Fermi statistics, the extra-attraction is lost \cite{Vijande:2013qr}.
The role of Fermi statistics is not mentioned in \cite{Bai:2016int}, and hence we believe that they predict binding for a $(QQ'\bar Q{\bar Q}')$ fictitious system with $Q\neq Q'$ but both having the same mass as the $b$ quark. This is similar to our finding in \cite{Vijande:2007ix}.

In a purely linear model, with kinetic energy and string potential \eqref{eq:tetraquark}, once $(QQ'\bar Q{\bar Q}')$ is stable, this remains true against any change of the string constant $\lambda$ and quark mass $M$. As shown in Appendix \ref{app:A}, the knowledge of the binding of  $(QQ'\bar Q{\bar Q}')$ below its threshold provides a minimum extension of the stability domain of $(bc\bar b\bar c)$ around the point $m_b=m_c$, in the plane of the masses. Beyond this extension, the question can be raised of the metastability of possible $(bc\bar b\bar c)$  resonances lying below their highest threshold  $(b\bar c)+(c \bar b)$ but above the $(b\bar b)+(c\bar  c)$ one. 

% If one extrapolates to lower quark masses, this property explain why $(c q\bar c\bar q)$ states are near or below their $(c\bar q)+\mathrm{c.c.}$ threshold, before the many corrections: spin effects, chiral dynamics, long-range Yukawa forces, etc. 

%
\section{Chromomagnetic interaction}\label{se:CMI}
This is the best known contribution to the energy of hadrons.  Within the 70s, it provided with 
a convincing explanation of the hyperfine splittings~\cite{DeRujula:1975ge}, and the first speculation 
on multiquarks bound mainly by chromomagnetic forces~\cite{Jaffe:1976yi}. 

In its most schematic version, it reads
\begin{equation}
 \label{eq:CM}
 V_\text{CM}=- C\,\sum_{i<j} \frac{\llss{i}{j}}{m_i\,m_j}\,\delta^{(3)}(\vec r_{ij})~.
\end{equation}

The remarkable feature of this Hamiltonian, stressed in \cite{Jaffe:1976yi}, is that the sum of strength, $C\langle \sum_{i<j} \llss{i}{j}\rangle$, might be larger than the cumulated values in the hadrons constituting the threshold, and thus lead to binding. This contrasts with the rule $\sum g_{ij}=2$ in the additive chromoelectric model of \eqref{eq:colour-add}, which reflects the overall color neutrality. 

In the first study of the $H=(uuddss)$ dibaryon, several simplifying approximations were made: SU(3) symmetry, a first order treatment of $V_\text{CM}$ (otherwise the contact term should be regularized, as done in several more recent studies), and moreover, the assumption that the short-range correlation factor $C_{ij}=C\,\langle \delta^{(3)}(\vec r_{ij}) \rangle$ assumes the \emph{same} value in the hexaquark and in ordinary baryons. These hypotheses were shown to artificially enhance the possibility of binding, see, e.g., \cite{Oka:1983ku,Rosner:1985yh}, etc.

Unfortunately, the computation of the short-range correlations $C_{ij}$  remains almost as difficult as it was in 1977, and its value in multiquarks is most often inferred from ordinary hadrons, as in \cite{Karliner:2016zzc,Hogaasen:2005jv}. 

Another warning is that in Jaffe's paper on $H$, as in some subsequent papers \cite{Hogaasen:2005jv,Wu:2016vtq}, only the chromomagnetic part is taken into account. This means that somehow, it is implicitly assumed that there is a draw between the multiquark and its threshold when only the kinetic and chromoelectric parts are included. This is not always the case \cite{Vij16,Oka:1983ku}.

\section{Outlook}
Let us summarize the most relevant issues concerning multiquark hunting:
\begin{itemize}
\item 
On the experimental side, identifying fully-heavy tetraquark bound states or resonances would be very welcome, to probe the confinement dynamics and confront different approaches to QCD.
\item 
The double hidden-flavor configurations could perhaps be more easily accessible with most detectors, with well-identified real or virtual quarkonia in the final state. 
\item 
However, many interesting issues deal with open flavor states, such as $(bb\bar c\bar c)$, and if a  $J/\psi$ trigger remains a powerful tool  in experimental set-ups,  it should not become an addiction that restricts the investigations in the space of flavors.  
Before $(bb\bar c\bar c)$, lighter states remain to be revealed, and a simultaneous search of double-charm baryons and double-charm tetraquarks $(cc\bar q\bar q)$ is clearly a priority.
\item 
On the theoretical side, there is a very rich physics in the light-quark sector, that potential models hardly take into account: more important role of chromomagnetic effects, chiral dynamics, long-range Yukawa forces \cite{Tornqvist:1993ng}, relativistic effects \cite{Heupel:2012ua}, etc. 
The fully-heavy tetraquarks give a chance to probe whether potential models, which are very successful for quarkonia, can be extended for higher configurations.
\item 
For any choice of the four-quark dynamics, calculating the energy and wavefunction of the tetraquark system is far from obvious.
In a four-body system, there is a sharp competition between building a collective configuration and splitting into two clusters. 
\item
The chromoelectric model with additive potentials bear many similarities with few-charge systems in atomic physics, especially for the mass-dependence of the binding energies. It differs, however, in the case of equal masses: while the positronium molecule $\mathrm{Ps}_2$ is weakly bound, $(cc\bar c\bar c)$ remains unbound with additive chromoelectric potentials.
\item 
The adiabatic version of the string model of confinement, though very appealing, holds for non-identical quarks, and provides more attraction that the usual pairwise models.  In particular, the configuration  $(bc\bar b \bar c)$ makes full use of the string dynamics.
 \item The chromomagnetic interaction is just one piece of the story. If one accounts for both chromoelectric and chromomagnetic terms, one hardly avoids a careful treatment of the four-body problem, as each term suggests a different type of clustering and/or color coupling.
 \item The diquark-antidiquark models require additional assumptions. If one compares the diquark picture to a standard quark model, one realizes that the former artificially lowers the chromoelectric contribution to the multiquark energy. Some further hypotheses are sometimes made about the chromomagnetic interaction inside the diquarks or between diquarks \cite{Maiani:2014aja}.
 %especially when the chromomagnetic is switched on or off \textsl{ad libidum} inside or outside each diquark.
 
%  \item The case of flavour asymmetric configurations remains open. In particular, a simultaneous search for double-charm baryons and double-charm tetraquarks should be seriously undertaken. 
% % 
%  \item The  $J/\psi$ trigger is a powerful tool  in experimental set-ups, but it should not become an addiction that restricts the investigation to the sectors with hidden heavy flavour. 
% % 
%  \item To sum up, careful four-body calculations based on simple models give $(cc\bar c\bar c)$ and $(bb\bar b\bar b)$ \emph{above} their dissociation threshold, while the case of $(bc\bar b\bar c)$ is more favourable. Of course, the conclusions might be different with a more elaborate dynamical framework, as in \cite{Heupel:2012ua}. 
%
\item 
The stability and instability patterns observed for $(Q_1Q_2{\bar Q}_3{\bar Q}_4)$ tetraquarks can give some guidance for higher configurations, for instance pentaquarks $(Q_1Q_2 (QQ'){\bar Q}_4)$ or hexaquarks $(Q_1Q_2 (QQ')(QQ'))$, starting from the limit where one or two sets of quarks are strongly correlated. It is our intend to extend our investigations to fully-heavy pentaquarks and hexaquarks. 
\end{itemize}

\subsection*{Acknowledgments}
This work has been partially funded  by Ministerio de Educaci\'on y
Ciencia and EU FEDER under Contracts No. FPA2013-47443,
FPA2015-69714-REDT and FPA2016-77177, by Junta de Castilla y Le\'on
under Contract No. SA041U16, and by Generalitat Valenciana
PrometeoII/2014/066.

\appendix
\section{Metastability of $C$-symmetric, unequal-mass configurations}\label{app:A}
Here we show that the stability of an equal-mass configuration $(QQ'\bar Q \bar{Q}')$, where the quarks $Q$ and $Q'$ have the same mass $M$ but are distinguishable, implies the stability of any $(Q_1 Q_2{\bar Q}_1{\bar Q}_2)$ with respect to its threshold $(Q_1\bar Q_2)+\mathrm{c.c.}$, provided the Hamiltonian is flavor-independent and the masses obey $2/M=1/M_1+1/M_2$.  

The reasoning is very similar to  the proof that the stability of $\mathrm{Ps}_2$ implies that of $\mathrm{H}_2$ \cite{ARV}. In an obvious notation, the $(Q_1 Q_2{\bar Q}_1{\bar Q}_2)$ Hamiltonian reads
\begin{equation}\label{eq:A1}
 H=X_1\,(\vec{p}_1^2+\vec{p}_3^2)+ X_2\,(\vec{p}_2^2+\vec{p}_4^2)+V~,
\end{equation}
where $X_1=1/(2\,M_1)$ and $X_2=1/(2\,M_2)$. It can be decomposed into
\begin{equation}
\begin{aligned}\label{eq:A2}
H&=H_s+H_a~,\\
H_s&=\frac{X_1+X_2}{2}\,(\vec{p}_1^2+\vec{p}_3^2+\vec{p}_2^2+\vec{p}_4^2)+V~,\\
H_a&=\frac{X_1-X_2}{2}\,(\vec{p}_1^2+\vec{p}_3^2-\vec{p}_2^2-\vec{p}_4^2)~,
\end{aligned}
\end{equation}
where $H_s$ is symmetric and $H_a$ antisymmetric with respect to $C$-conjugation, i.e., simultaneous $1\leftrightarrow 2$ and $3\leftrightarrow 4$. The asymmetric part $H_a$ lowers the energy of $H$ as compared to the ground-state energy of the symmetric part $H_s$ alone, which is the Hamiltonian of $(QQ'\bar Q \bar{Q}')$.  Moreover, the threshold of $H_s$ and the $(Q_1\bar Q_2)+\mathrm{c.c.}$ threshold of $H$ have the same energy, since they are governed by the same inverse reduced mass. Thus, if the brackets denote the lowest energy 
%\leftrightarrow
\begin{equation}\label{eq:A3}
 [QQ'\bar Q{\bar Q}']<2\,[Q\bar Q] \ \Rightarrow\ 
 [Q_1Q_2{\bar Q}_1{\bar Q}_2]<2\,[Q_1{\bar Q}_2]~.
\end{equation}
For a purely linear potential, the stability of $(QQ'\bar Q{\bar Q}')$, if true for some quark mass $M$, holds for any other mass, as a change of mass induces the same scaling factor for the mesons and for the tetraquark. For a more complicated interaction, \eqref{eq:A3} requires 
the condition $2/M=1/M_1+1/M_2$ on the inverse masses. 

The knowledge of the binding in the equal-mass case, measured by the parameter $\epsilon$ defined as
\begin{equation}\label{eq:A4}
[QQ'\bar Q{\bar Q}']=2\,[Q\bar Q] (1-\epsilon)~,
\end{equation}
indicates a minimal range of masses $M_1$ and $M_2$ around $M_1=M_2=M$ for which stability remains. 
In the case of a purely linear interaction, it is given by 
\begin{equation}
\begin{aligned}\label{eq:A5}
1/M_1+1/M_2&=2/M~\\
M_1^{-1/3}+M_2^{-1/3}&\ge 2\,(1-\epsilon)\,M^{-1/3}~
\end{aligned}
\end{equation}
which can be solved in closed form. For instance, with an energy $[QQ'\bar Q{\bar Q}']\simeq 4.639$ \cite{Vijande:2007ix} for a unit-mass tetraquark bound by string-potential of unit strength, to be compared to a threshold $2\,[Q\bar Q]=4.676$ (twice the negative the first zero of the Airy function), the stability of $(Q_1Q_2{\bar Q}_1{\bar Q}_2)$ is guaranteed for at least $1/1.72<M_1/M_2<1.72$.

In the case of a more general flavor-independent interaction, the second equation of \eqref{eq:A5} should be replaced by an exact estimate of the two-meson lowest threshold as a function of $M_1$ and $M_2$.

% \bibliographystyle{unsrt}
% \bibliography{multi}

\begin{thebibliography}{10}

\bibitem{2015Natur.523R.267C}
M.~{Chalmers}.
\newblock {Forsaken pentaquark particle spotted at CERN}.
\newblock {\em 523:267--268, July 2015.}

\bibitem{LHCB-CC}
LHCb finds tetraquark candidates, CERN Courier, August 12, 2016,
  http://cerncourier.com/cws/article/cern/65798.

\bibitem{Karliner:2016zzc}
Marek Karliner, Jonathan~L. Rosner, and Shmuel Nussinov.
\newblock {Production and Decay of (Q Q Qbar Qbar) States}.
\newblock 2016.

\bibitem{Bai:2016int}
Yang Bai, Sida Lu, and James Osborne.
\newblock {Beauty-full Tetraquarks}.
\newblock 2016.

\bibitem{Wu:2016vtq}
Jing Wu, Yan-Rui Liu, Kan Chen, Xiang Liu, and Shi-Lin Zhu.
\newblock {Heavy-flavored tetraquark states with the $QQ\bar{Q}\bar{Q}$
  configuration}.
\newblock 2016.

\bibitem{Wang:2017jtz}
Zhi-Gang Wang.
\newblock {Analysis of the $QQ\bar{Q}\bar{Q}$ tetraquark states with QCD sum
  rules}.
\newblock 2017.

\bibitem{Iwasaki01081975}
Yoichi Iwasaki.
\newblock A possible model for new resonances: Exotics and hidden charm.
\newblock {\em Progress of Theoretical Physics}, 54(2):492--503, 1975.

\bibitem{Carlson:1991zt}
J.~Carlson and V.~R. Pandharipande.
\newblock Absence of exotic hadrons in flux tube quark models.
\newblock {\em Phys. Rev.}, D43:1652--1658, 1991.

\bibitem{Lloyd:2003yc}
Richard~J. Lloyd and James~P. Vary.
\newblock {All charm tetraquarks}.
\newblock {\em Phys. Rev.}, D70:014009, 2004.

\bibitem{Zouzou:1986qh}
S.~Zouzou, B.~Silvestre-Brac, C.~Gignoux, and J.~M. Richard.
\newblock {Four quark bound states}.
\newblock {\em Z. Phys.}, C30:457, 1986.

\bibitem{Vijande:2009kj}
J.~Vijande, A.~Valcarce, and N.~Barnea.
\newblock {Exotic meson-meson molecules and compact four--quark states}.
\newblock {\em Phys. Rev.}, D79:074010, 2009.

\bibitem{Lipkin:1986dw}
Harry~J. Lipkin.
\newblock A model independent approach to multi - quark bound states.
\newblock {\em Phys. Lett.}, B172:242, 1986.

\bibitem{Lipkin:2007cg}
Harry~J. Lipkin.
\newblock {Exotic hadrons in the constituent quark model}.
\newblock {\em Prog. Theor. Phys. Suppl.}, 168:15--22, 2007.

\bibitem{1946NYASA..48..219W}
J.~A. {Wheeler}.
\newblock {Polyelectrons}.
\newblock {\em Annals of the New York Academy of Sciences}, 48:219, December
  1946.

\bibitem{PhysRev.70.90}
Aadne Ore.
\newblock Binding energy of polyelectrons.
\newblock {\em Phys. Rev.}, 70(1-2):90, Jul 1946.

\bibitem{PhysRev.71.493}
Egil~A. Hylleraas and Aadne Ore.
\newblock Binding energy of the positronium molecule.
\newblock {\em Phys. Rev.}, 71(8):493--496, Apr 1947.

\bibitem{ARV}
E.~A.~G. Armour, J.~M. Richard, and K.~Varga.
\newblock Stability of few-charge systems in quantum mechanics.
\newblock {\em Phys. Rep.}, 413:1--90, 2004.

\bibitem{2004FBS....34...63F}
P.~{Froelich}, B.~{Zygelman}, A.~{Saenz}, S.~{Jonsell}, S.~{Eriksson}, and
  A.~{Dalgarno}.
\newblock {Hydrogen-Antihydrogen Molecule and Its Properties}.
\newblock {\em Few-Body Systems}, 34:63--72, 2004.

\bibitem{Eic75}
E.~Eichten, K.~Gottfried, T.~Kinoshita, K.~Kogut, K.D. Lane, and T.-M. Yan.
\newblock {\em Phys. Rev. Lett}, 34:369, 1975.
\newblock [Erratum: Phys. Rev. Lett. 36, 1276(1976)].

\bibitem{Quigg:1979vr}
C.~Quigg and Jonathan~L. Rosner.
\newblock {Quantum Mechanics with Applications to Quarkonium}.
\newblock {\em Phys.Rept.}, 56:167--235, 1979.

\bibitem{Martin:1980rm}
Andre Martin.
\newblock {A simultaneous fit of $b\bar b$, $c \bar c$, $s \bar s$, (BCS pairs)
  and $c\bar s$ spectra}.
\newblock {\em Phys. Lett.}, B100:511--514, 1981.

\bibitem{Janc:2004qn}
D.~Janc and M.~Rosina.
\newblock {The $T_{cc}= D D^*$ molecular state}.
\newblock {\em Few Body Syst.}, 35:175--196, 2004.

\bibitem{Vijande:2007rf}
J.~Vijande, E.~Weissman, A.~Valcarce, and N.~Barnea.
\newblock {Are there compact heavy four-quark bound states?}
\newblock {\em Phys. Rev.}, D76:094027, 2007.

\bibitem{Olive:2016xmw}
C.~Patrignani et~al.
\newblock {Review of Particle Physics}.
\newblock {\em Chin. Phys.}, C40(10):100001, 2016.

\bibitem{Richard:2010jh}
Jean-Marc Richard.
\newblock {Non-Abelian dynamics and heavy multiquarks, Steiner-tree confinement
  in hadron spectroscopy}.
\newblock {\em Few Body Syst.}, 50:137--143, 2011.

\bibitem{Richard:2016eis}
Jean-Marc Richard.
\newblock {Exotic hadrons: review and perspectives}.
\newblock {\em Few Body Syst.}, 57(12):1185--1212, 2016.
\newblock {Special issue for the 30th anniversary of Few-Body Systems}.

\bibitem{Vij16}
J.~Vijande, A.~Valcarce, J.-M Richard, and P.~Sorba.
\newblock {\em Phys. Rev.}, D94:034038, 2016.

\bibitem{Vij09}
J.~Vijande and A.~Valcarce.
\newblock {\em Phys. Rev.}, C80:035204, 2009.

\bibitem{Bar07}
N.~Barnea, J.~Vijande, and A.~Valcarce.
\newblock {\em Phys. Rev.}, D73:054004, 2006.

\bibitem{Nussinov:1999sx}
Shmuel Nussinov and Melissa~A. Lampert.
\newblock Qcd inequalities.
\newblock {\em Phys. Rept.}, 362:193--301, 2002.

\bibitem{Bertlmann:1979zs}
R.~A. Bertlmann and Andre Martin.
\newblock {I\lowercase{NEQUALITIES ON HEAVY QUARK - ANTIQUARK SYSTEMS}}.
\newblock {\em Nucl. Phys.}, B168:111--136, 1980.

\bibitem{bransden2003physics}
B.H. Bransden and C.J. Joachain.
\newblock {\em Physics of Atoms and Molecules}.
\newblock Pearson Education. Prentice Hall, 2003.

\bibitem{reif2009fundamentals}
F.~Reif.
\newblock {\em Fundamentals of Statistical and Thermal Physics:}.
\newblock Waveland Press, 2009.

\bibitem{Braaten:2014qka}
Eric Braaten, Christian Langmack, and D.~Hudson Smith.
\newblock {Born-Oppenheimer Approximation for the $XYZ$ Mesons}.
\newblock {\em Phys. Rev.}, D90(1):014044, 2014.

\bibitem{Semay:1994ht}
C.~Semay and B.~Silvestre-Brac.
\newblock {Diquonia and potential models}.
\newblock {\em Z. Phys.}, C61:271--275, 1994.

\bibitem{Car12}
T.F. Caram\'es, A.~Valcarce, and J.~Vijande.
\newblock {\em Phys. Lett.}, B709:358, 2012.

\bibitem{Dosch:1975gf}
Hans~Gunter Dosch and V.~F. Muller.
\newblock On composite hadrons in nonabelian lattice gauge theories.
\newblock {\em Nucl. Phys.}, B116:470, 1976.

\bibitem{Lenz:1985jk}
F.~Lenz, J.~T. Londergan, E.~J. Moniz, R.~Rosenfelder, M.~Stingl, and
  K.~Yazaki.
\newblock {Quark Confinement and Hadronic Interactions}.
\newblock {\em Annals Phys.}, 170:65, 1986.

\bibitem{Vijande:2007ix}
J.~Vijande, A.~Valcarce, and J.~M. Richard.
\newblock {Stability of multiquarks in a simple string model}.
\newblock {\em Phys. Rev.}, D76:114013, 2007.

\bibitem{Vijande:2013qr}
Javier Vijande, Alfredo Valcarce, and Jean-Marc Richard.
\newblock {Adiabaticity and color mixing in tetraquark spectroscopy}.
\newblock {\em Phys. Rev.}, D87(3):034040, 2013.

\bibitem{DeRujula:1975ge}
A.~De~R{\'u}jula, Howard Georgi, and S.~L. Glashow.
\newblock H\lowercase{ADRON MASSES IN A GAUGE THEORY}.
\newblock {\em Phys. Rev.}, D12:147--162, 1975.

\bibitem{Jaffe:1976yi}
Robert~L. Jaffe.
\newblock {Perhaps a Stable Dihyperon}.
\newblock {\em Phys. Rev. Lett.}, 38:195--198, 1977.
\newblock [Erratum: Phys. Rev. Lett.38,617(1977)].

\bibitem{Oka:1983ku}
M.~Oka, K.~Shimizu, and K.~Yazaki.
\newblock {The Dihyperon State in the Quark Cluster Model}.
\newblock {\em Phys. Lett.}, B130:365, 1983.

\bibitem{Rosner:1985yh}
Jonathan~L. Rosner.
\newblock {SU(3) Breaking and the $H$ Dibaryon}.
\newblock {\em Phys. Rev.}, D33:2043, 1986.

\bibitem{Hogaasen:2005jv}
H.~H\o{}g\aa{}sen, J.-M. Richard, and P.~Sorba.
\newblock {A chromomagnetic mechanism for the X(3872) resonance}.
\newblock {\em Phys. Rev.}, D73:054013, 2006.

\bibitem{Tornqvist:1993ng}
Nils~A. T{\"o}rnqvist.
\newblock {From the deuteron to deusons, an analysis of deuteron - like meson
  meson bound states}.
\newblock {\em Z. Phys.}, C61:525--537, 1994.

\bibitem{Heupel:2012ua}
Walter Heupel, Gernot Eichmann, and Christian~S. Fischer.
\newblock {Tetraquark Bound States in a Bethe-Salpeter Approach}.
\newblock {\em Phys. Lett.}, B718:545--549, 2012.

\bibitem{Maiani:2014aja}
L.~Maiani, F.~Piccinini, A.~D. Polosa, and V.~Riquer.
\newblock {The Z(4430) and a New Paradigm for Spin Interactions in
  Tetraquarks}.
\newblock {\em Phys. Rev.}, D89:114010, 2014.

\end{thebibliography}
\end{document}